\documentstyle[pra,aps,multicol,psfig]{revtex}

\def\nm{\sigma}  

\def\logA{\mbox{ln}}

\def\multicolsbeg#1{\begin{multicols}{#1}}
\def\multicolsend{\end{multicols}}

\begin{document}
\title{Partitioning optical solitons for generating entangled
light beams}
\author{
  Eduard Schmidt,
  Ludwig Kn\"oll, and
  Dirk--Gunnar Welsch
}
\address{Friedrich-Schiller-Universit\"{a}t Jena,
  Theoretisch-Physikalisches Institut\\
  Max-Wien-Platz 1, D-07743 Jena, Germany}
\date{\today}
\maketitle

\begin{abstract}
It is shown that bipartition of optical solitons 
can be used to generate entangled light beams.
The achievable amount of entanglement 
can be substantially larger for $N$-bound 
solitons ($N$ $\!=$ $\!2,3$) than for the
fundamental soliton ($N$ $\!=$ $\!1$). An analysis
of the mode structure of the entangled beams shows
that just $N$ modes are essentially entangled.
In particular, partitioning of the fundamental soliton 
effectively produces $2$-mode squeezed light.
\end{abstract}

\vspace*{1cm}
\frenchspacing
\multicolsbeg{2}

Quantum coherent dynamics of atomic and photonic systems 
has been a subject of increasing interest stimulated by 
the rapidly developing field of quantum information processing
and quantum computing 
\cite{DeutschD85,BennettCH95,EkertA96,QuantInfSet01,SteaneA98}.
Intrinsic parallelism of quantum state evolution
renders it possible to overcome the problem of
exponential time required by a classical computer
for solving complex problems such as integer factorization 
\cite{ShorPW94,EkertA95}, discrete logarithm \cite{ShorPW94},
and search \cite{GroverLK97}. High sensitivity of quantum coherent 
dynamics to eavesdropping can be used for developing secure encryption 
protocols \cite{BennettCH84} realized in fiber optical systems 
\cite{ZbindenH98}. 

One of the most remarkable features of quantum coherence
is entanglement, on which, e.g., quantum teleportation
is based \cite{Zeilinger}. Entanglement can be regarded as 
being the nonclassical contribution to the overall correlation
between two parts of a system (see, e.g., \cite{BennettCH96,VedralV97}).
Typically, discrete systems that are built up by qubits have 
been considered. Concepts that use continuous quantum variables 
have recently attracted increasing publicity     
\cite{VaidmanL94,BraunsteinSL98,FurusawaA98,LloydS98,%
BraunsteinSL98a,LloydS99}. An illustrative example is 
the teleportation of coherent states by means of 
entangled light of the squeezed-vacuum type \cite{FurusawaA98}.

A way to generate entangled light is to use two 
fields each of which is prepared in some 
nonclassical state and combine them at a linear four-port 
device like a beam splitter or a coupler.
In particular, the entangled squeezed light used
in \cite{FurusawaA98} is produced by combining two 
light beams independently squeezed via $\chi^{(2)}$ 
parametric processes. The possibility of
realizing entangled pulse sources by combining
two separately squeezed optical solitons
are discussed in \cite{LeuchsG99}.
Alternatively, a nonlinear coupling between two fields can 
prepare them in an entangled (continuous-variable) state.  
For example, the $\chi^{(2)}$ process 
of parametric down-conversion has been extensively 
used as a source of entangled beams and pulses
(see, e.g., \cite{JoobeurA96,KellerTE97}).
In this letter we show that bipartition of 
(fundamental and $N$-bound) optical solitons
formed in a medium with a Kerr-type $\chi^{(3)}$ nonlinearity
yields entangled light fields, because of the internal 
quantum correlations of the solitons. 


{F}rom classical optics it is well known that
the Kerr nonlinearity can compensate for dispersion-assisted 
pulse spreading or diffraction-assisted beam broadening
(see, e.g., \cite{HasegawaA89,ReynaudF92}).
In both cases, the evolution of the (complex)
field amplitude $a(x,t)$ can be described 
by the nonlinear Schr\"odinger equation (NSE) 
\begin{eqnarray}
&&i \partial _{t} a(x,t) = 
-{\textstyle\frac{1}{2}}\omega ^{(2)}
  \partial_{xx} a(x,t) + \chi|a(x,t)|^2 a(x,t),
\label{eq.NSE}
\end{eqnarray}
($t$, propagation variable; $x$, ``transverse'' coordinate;
$\omega^{(2)}$, second-order dispersion or diffraction constant;
${\chi }$, nonlinear-coupling constant).
Bright temporal solitons can be formed either
in focusing media with anomalous dispersion
(\mbox{$\chi$ $\!<$ $0$}, \mbox{$\omega^{(2)}$ $\!>$ $\!0$}) or
in defocusing media with normal dispersion
(\mbox{$\chi$ $\!>$ $\!0$}, \mbox{$\omega^{(2)}$ $\!<$ $\!0$}).
Spatial solitons always require a focusing nonlinearity.
Instead of working in $x$-space, we can also turn to the $\omega$-space,
\begin{equation}
\underline{a}(\omega,t)=\frac{1}{\sqrt{2\pi}}
\int_{-\infty}^{\infty} dx\, a(x,t) e^{i \omega x} .
\label{eq.aw}
\end{equation}
In what follows the coordinates in the $x$- and $\omega$-domains 
are scaled by the initial soliton width $x_0$ and the spectral width
$\omega_0$ $\!=$ $\!1/x_0$ respectively. Propagation distances are 
measured in dispersion (diffraction) lengths \mbox{$t_{\rm d}$ 
$\!=$ $\!x_0^2/|\omega^{(2)}|$}. The NSE (\ref{eq.NSE})
can be solved by the inverse scattering method \cite{ZakharovVE72}.
In particular, the bell-shaped fundamental soliton 
is extremely stable and can propagate over distances
far beyond the limit given by the 
dispersion- (diffraction-) assisted spreading.
Scaling the power of the fundamental soliton by a factor of $N^2$,
$N$ $\!=$ $\!2,3,\ldots$, yields so-called $N$-bound solitons 
\cite{SatsumaJ74}.
The two classes of solutions of the NSE have been realized 
experimentally (see, e.g., \cite{MollenauerLF80,BarthelemyA85,ManeufS88}).

In quantum theory, the complex field amplitude $a(x,t)$ and
$a^\ast(x,t)$ become field operator $\hat{a}(x,t)$ and
$\hat{a}^\dagger(x,t)$, respectively, with 
\begin{eqnarray}
&&\left[ \hat{a}(x,t),\hat{a}^{\dagger }(x^{\prime },t)\right] =
\delta (x-x^{\prime }),
\label{eq.comm.rel}
\end{eqnarray}
and the NSE becomes an operator-valued equation 
\cite{LaiY89,MecozziA98}. 
In order to solve the quantum mechanical problem, it is commonly 
assumed that the initial state is a multimode coherent state such that
$\langle\hat{a}\rangle$ $\!\sim$ $\!N \mbox{sech}(x/x_0)$
is the initial shape of the classical $N$-bound-soliton solution.
For not too large propagation distances the state then evolves  
into a multimode squeezed state of Gaussian type, which
can be calculated numerically within the framework of 
appropriate discretization \cite{SchmidtE99}.

\begin{figure}[tbp]
\centerline{
\psfig{file=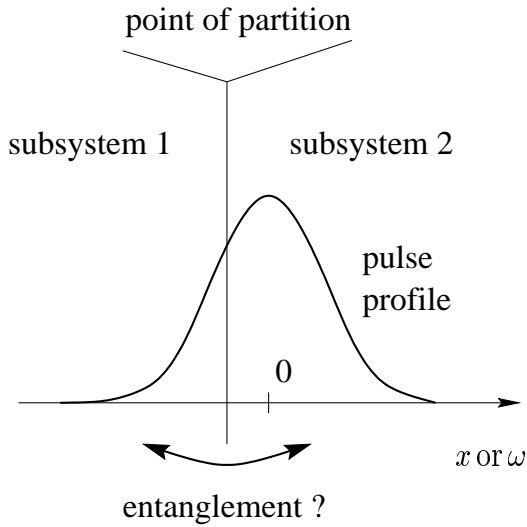,width=2.8in}
}
\vspace*{0.5cm}
\caption{
Bipartition of optical solitons. Applying appropriate
filtering techniques (see, e.g., \protect\cite{MecozziA98,%
FribergSR96,SpaelterS97}), it can be realized in the $x$-domain
by partition of the near field of spatial solitons 
and in the $\omega$-domain by partition of the far field of 
spatial solitons or spectral partition of temporal solitons.
}
\label{fig1}
\end{figure}

The nonlinear soliton motion leads to internal quantum 
correlations, which have been studied both experimentally 
\cite{SpaelterS98} and theoretically 
\cite{SchmidtE99,MecozziA97,LevandovskyD99,SchmidtE2000}).
{F}rom the results the question is suggested of whether or not it is
possible to produce two entangled light beams by
appropriate partitioning of quantum solitons. 
Let us consider bipartition as sketched 
in Fig.~\ref{fig1}, which may be realized experimentally using
spectral filtering \cite{FribergSR96,SpaelterS97}
or spatial filtering \cite{MecozziA98}.
Since the soliton quantum state remains pure during 
the propagation, the entanglement $E$ of
the bipartite system is given by the von Neumann entropy
of a subsystem \cite{BennettCH96}
i.e.,
\begin{equation}
\label{eq.S1}
E = S_1 = - {\rm Tr}_1(\hat{\rho}_1 \logA\hat\rho_1)
= S_2 = - {\rm Tr}_2(\hat{\rho}_2 \logA\hat\rho_2),
\end{equation}
where $\hat{\rho}_i$ is the density operator 
of the $i$th subsystem, and ${\rm Tr}_i$ means the trace 
with regard to the $i$th subsystem \mbox{($i$ $\!=$ $\!1,2$)}.

\vspace*{0.5cm}

In order to calculate the entropy of a subsystem prepared 
in a Gaussian state, let us consider the $x$-domain and 
assume that the subsystem extends over a region $\Xi$
(the calculation in the $\omega$-domain is analogous). 
We introduce new bosonic operators  
\begin{equation}
\hat{b}_k = \int\nolimits_{\Xi} dx\,
      \left[
       \mu_{k}(x)\,\Delta\hat{a}(x)
       +\nu_{k}(x)\,\Delta\hat{a}^\dagger\!(x)
      \right]
\label{eq.bk}
\end{equation}
[$\Delta\hat{a}(x)$ $\!=$ $\!\hat{a}(x)$ $\!-$ $\!\langle\hat{a}(x)\rangle$]
and impose on the $\mu_{k}(x)$ and $\nu_{k}(x)$ the condition that 
\begin{eqnarray}
 \langle\hat{b}_k^\dagger\hat{b}_{k'}^{}\rangle=\nm_k\delta_{kk'}, \quad
 \langle\hat{b}_k        \hat{b}_{k'}\rangle=0.
\label{eq.bb}
\end{eqnarray}
For notational convenience, we have omitted the argument $t$.
{F}rom Eqs.~(\ref{eq.bk}) and (\ref{eq.bb}) it then follows that the 
parameters $\nm_k$ and the functions $\mu_{k}(x)$ and $\nu_{k}(x)$ 
solve the equations
\begin{eqnarray}
\lefteqn{
  \left[\!
    \begin{array}{c}
      \left(
        \nm_k+{\textstyle\frac{1}{2}}
      \right)
      \delta_{kk'}\\
      0
    \end{array}
  \!\right]
=
  \int\nolimits_{\Xi} dx
  \int\nolimits_{\Xi} dx'
  \left[\!
    \begin{array}{cc}
      \mu_k(x)     & \nu_k(x)\\
      \nu_k^{*}(x) & \mu_k^{*}(x)
    \end{array}
  \!\right]
}
\nonumber\\[.5ex]&&\hspace{6ex}
  \times
  \left[\!
    \begin{array}{cc}
      C(x,x')^{*}&B(x,x')\\
      B(x,x')^{*}&C(x,x')
    \end{array}
  \!\right]
  \left[\!
    \begin{array}{c}
      \mu_{k'}^{*}(x')\\
      \nu_{k'}^{*}(x')
    \end{array}
  \!\right],
\label{eq.integr}
\end{eqnarray}
where
\begin{eqnarray}
&&\quad\quad
B(x,x')=\langle\Delta\hat{a}(x)\Delta\hat{a}(x')\rangle,
\label{eq.Bxx}
\\&&\quad\quad
C(x,x')=\langle\Delta\hat{a}^\dagger\!(x)\Delta\hat{a}(x')\rangle
+{\textstyle\frac{1}{2}}\delta(x-x').
\label{eq.Cxx}
\end{eqnarray}
Note that the functions 
$\mu_{k}(x)$ and $\nu_{k}(x)$ must satisfy the conditions
\begin{eqnarray}
&\displaystyle
\int\nolimits_{\Xi} dx\,
     \left[\mu_{k}(x)\,\mu_{k'}^{*}(x)
       -\nu_{k}(x)\,\nu_{k'}^{*}(x)
     \right]=\delta_{kk'},
&
\label{eq.mmnn}
\\[.5ex]
&\displaystyle
  \int\nolimits_{\Xi}  dx\,
     \left[\mu_{k}(x)\,\nu_{k'}(x)
       -\nu_{k}(x)\,\mu_{k'}(x)
     \right]=0, 
&     
\label{eq.mnmn}
\end{eqnarray}
because of the commutation relations 
$[\hat{b}_k,\hat{b}^\dagger_{k'}]$ $\!=$ $\!\delta_{kk'}$ and
$[\hat{b}_k,\hat{b}_{k'}]$ $\!=$ $\!0$.
With regard to the new bosonic operators,
the Gaussian state (of the subsystem) is obviously a multimode thermal 
state, with $\nm_k$ being the mean photon number of the $k$th mode.
Hence, the entanglement $E$ is given by
\begin{equation}
    E = \sum\nolimits_{k} S_{\rm th}(\nm_k),
\label{eq.Sk}
\end{equation}
with
\begin{equation}
    S_{\rm th}(\nm)=\logA\!\left[\frac{(\nm+1)^{\nm+1}}{\nm^\nm}\right]
\label{eq.Sth}
\end{equation}
being the entropy of a single-mode thermal state.
Note that when the $\Xi$-region comprises the whole soliton
prepared in a pure Gaussian state, then all the $\sigma_k$ vanish 
and thus $E$ $\!=$ $\!0$, as expected.  

\begin{figure}[tbp]
\centerline{
\psfig{file=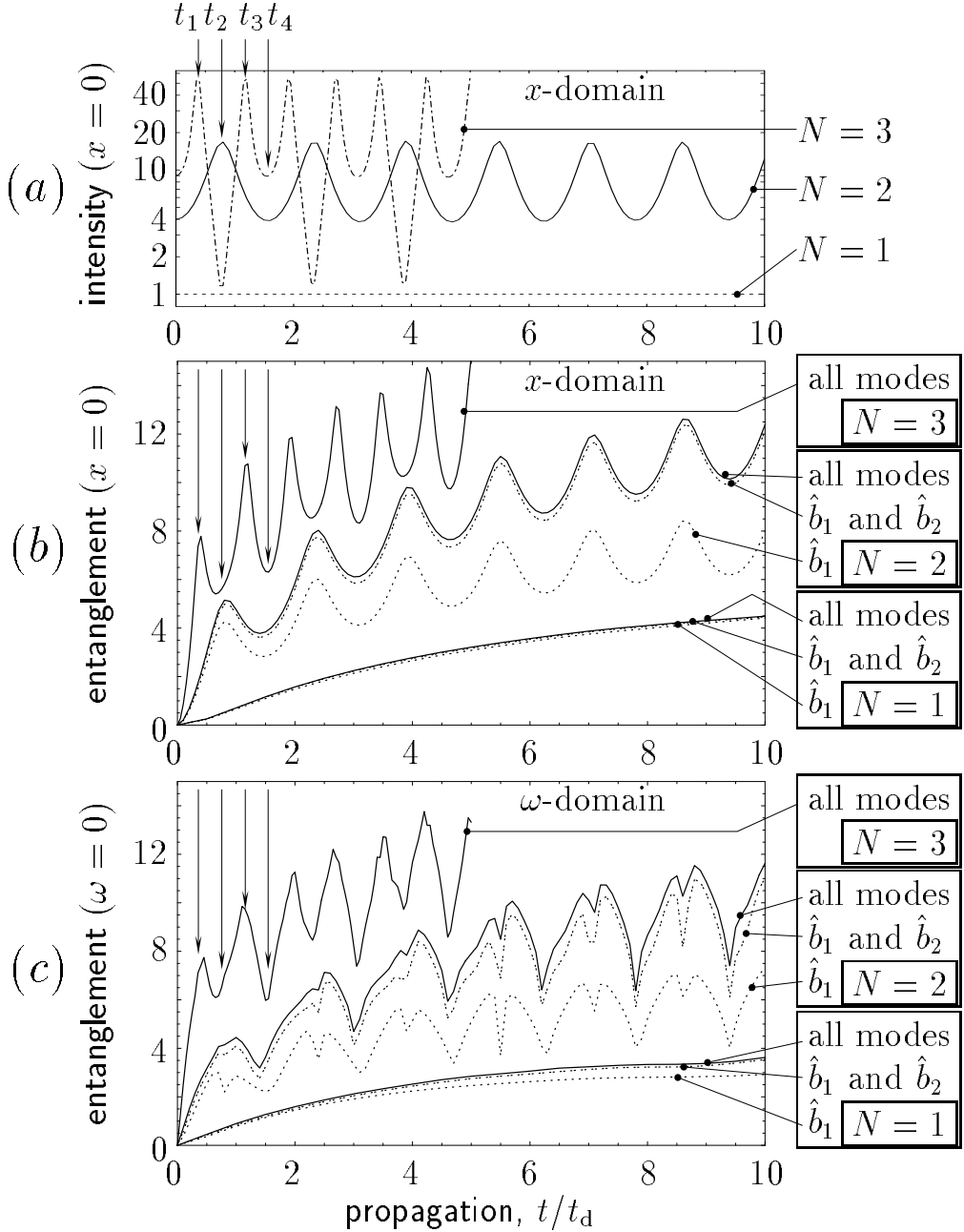,height=7in}
}
\vspace*{0.5cm}
\caption{
Dependence on the propagation distance of the
entanglement realized by symmetrical bipartition of the 
fundamental soliton ($N$ $\!=$ $\!1$) and $N$-bound solitons 
($N$ $\!=$ $\!2,3$) in the $x$-domain $(b)$ and the 
$\omega$-domain $(c)$. For comparison, the soliton mid-intensity
in the $x$-domain as a function of the propagation distance
is shown $(a)$. The contributions to the entanglement
of the relevant modes are shown for the fundamental soliton and
the $2$-bound soliton. The marked propagation distances are:
\mbox{$t_4$ $\!=$ $\!\textstyle\frac{\pi}{2}$ $\!t_{\rm d}$} (soliton period),
$t_2$ $\!=$ $\!\textstyle\frac{1}{2}$ $\!t_4$ 
(distance of compression for $2$-bound soliton),
$t_1$ $\!=$ $\!\textstyle\frac{1}{4}$ $\!t_4$ and 
$t_3$ $\!=$ $\!\textstyle\frac{3}{4}$ $\!t_4$ 
(distances of compression for $3$-bound soliton).
In the numerical calculations, which are performed on a grid of 
$256$ points with a discretization step 
\mbox{$\Delta x$ $\!=$ $\!0.05\,x_0$}, 
it is assumed that the total photon numbers of the solitons are 
$2N^2\bar{n}$, where $\bar{n}$ $\!=$ $\!10^{9}$
(for details, see \protect\cite{MecozziA98,SchmidtE99}).
}
\label{fig2}
\end{figure}

Figure \ref{fig2} presents examples of the entanglement 
obtained by symmetrical bipartition of the fundamental soliton 
and the 2- and 3-bound solitons in both the $x$-domain 
[solid lines in Fig.~\ref{fig2}$(b)$] and the $\omega$-domain 
[solid lines in Fig.~\ref{fig2}$(c)$] as a function of the 
propagation distance.
The figure clearly shows that highly entangled 
parties can be realized by bipartition of optical solitons 
produced in Kerr-like media.
The (numerical) calculations have shown that 
the maximum entanglement is typically observed
for symmetrical bipartition (i.e., at \mbox{$x$ $\!=$ $\!0$} 
or \mbox{$\omega$ $\!=$ $\!0$)}. 
For the system considered in the figure, an asymmetrical
bipartition of the $3$-bound soliton in the $x$-domain
at $|x|$ $\!\lesssim$ $\!0.4\,x_0$ can improve the 
entanglement by $5.3\%$ for certain propagation 
distances.

The entanglement realized by bipartition
of the fundamental soliton increases monotonously with the 
propagation distance. The maxima and minima of the entanglement
realized by bipartition in the $x$-domain of $2$- and $3$-bound solitons
follow quite exactly the periodic change between soliton compression
and expansion [compare Figs.~\ref{fig2}$(a)$ and \ref{fig2}$(b)$].
In the $\omega$-domain this correlation is less pronounced
[compare Figs.~\ref{fig2}$(a)$ and \ref{fig2}$(c)$].
It is worth noting that the amount of entanglement
that can be achieved increases with the order parameter $N$.

\begin{figure}[tbp]
\centerline{
\psfig{file=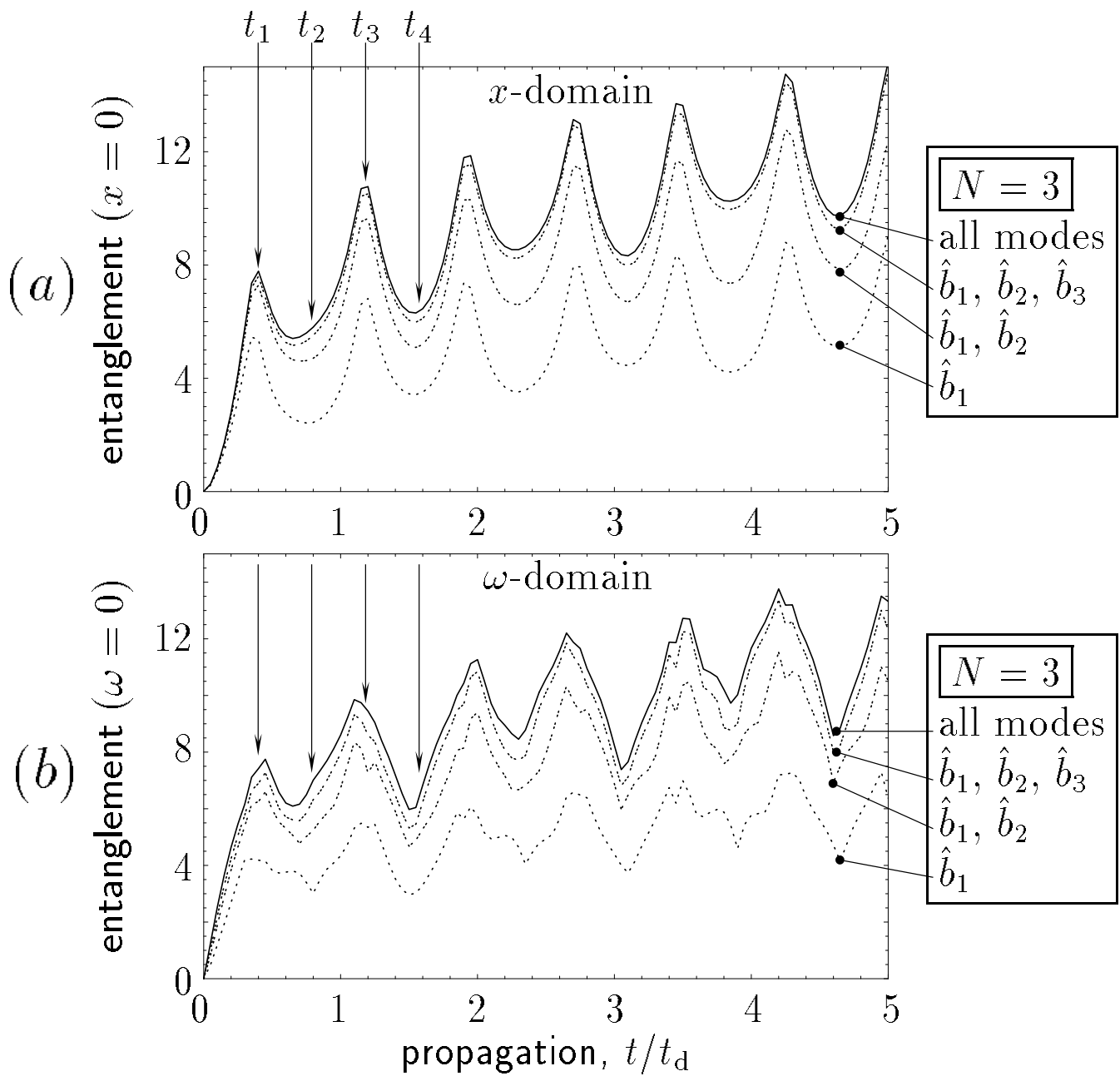,height=7in}
}
\vspace*{0.5cm}
\caption{
The contributions of the relevant modes to the entanglement
realized by symmetrical bipartition of the $3$-bound soliton
in the $x$-domain $(a)$ and the $\omega$-domain $(b)$.
The soliton is the same as in Fig.~\protect\ref{fig2}.
}
\label{fig3}
\end{figure}

Formally, the sum in Eq.~(\ref{eq.Sk}) runs over all modes.
In fact, only a few modes substantially contribute to the
entanglement, as it is seen from Figures \ref{fig2}
and \ref{fig3} (the modes are numbered such that the
single-mode contribution 
to the entanglement decreases with increasing
mode index). In particular, it is seen that 
the number of relevant modes is just given by the 
soliton-order parameter $N$. 
A possible reason for this fact can be seen in the
multicomponent structure of the classical $N$-soliton solution
as discussed in \cite{SchmidtE2000}. 

In summary, the results offer novel possibilities of
using optical solitons as sources of entangled light beams.
Although bipartition of solitons again yields multimode
objects, only a few modes are involved in the
entanglement. In particular, fundamental solitons are suited
for generation of entangled light like $2$-mode 
squeezed light, which is an alternative to the standard
$\chi^{(2)}$ sources.

\acknowledgments
This work was supported by the Deutsche Forschungsgemeinschaft.
We are grateful to
Toma{\v{s}}~Opatrn\'{y}
for valuable discussions.




\begin{thebibliography}{10}

\bibitem{DeutschD85}
D. Deutsch, Proc. R. Soc. London A {\bf 400},  97  (1985).

\bibitem{BennettCH95}
C.~H. Bennett, Phys. Today {\bf 48},  No. 10, 24  (1995).

\bibitem{EkertA96}
A. Ekert and R. Jozsa, Rev. Mod. Phys. {\bf 68},  733  (1998).

\bibitem{QuantInfSet01}
See the articles in the March 1998 issue of Phys. World.

\bibitem{SteaneA98}
A. Steane, Prog. Theor. Phys. {\bf 61},  117  (1998).

\bibitem{ShorPW94}
P.~W. Shor,  in {\em Proceedings of the 35th Symposium on the Foundations of
  Computer Science} (IEEE Computer Society Press, Los Alamitos, 1994), p.\ 124.

\bibitem{EkertA95}
A. Ekert and R. Jozsa, Shor's quantum algorithm for factorising numbers,
  preprint, Dept. of Mathematics and Statistics, Univ. of Plymouth, Plymouth,
  Devon, UK, 1995.

\bibitem{GroverLK97}
L.~K. Grover, Phys. Rev. A {\bf 79},  325  (1997).

\bibitem{BennettCH84}
C.~H. Bennett and G. Brassard, Proc. Int. Conf. Computer Systems and Signal
  Processing {\bf 175},  Bangalore  (1984).

\bibitem{ZbindenH98}
H. Zbinden, H. Bechmann-Pasquinucci, N. Gisin, and G. Ribordy, Appl. Phys. B
  {\bf 67},  743  (1998).

\bibitem{Zeilinger}
D. Bouwmeester, J.-W. Pan, M. Daniel, H. Weinfurter, and A. Zeilinger, 
Nature {\bf 390}, 575 (1997);
D. Boschi, S. Branca, F. DeMartini, L. Hardy, and S. Popescu, 
Phys. Rev. Lett. {80}, 1121 (1998).

\bibitem{BennettCH96}
C.~H. Bennett, D.~P. DiVincenzo, J.~A. Smolin, and W.~K. Wootters, Phys. Rev. A
  {\bf 54},  3824  (1996).

\bibitem{VedralV97}
V. Vedral, M.~B. Plenio, M.~A. Rippin, and P.~L. Knight, Phys. Rev. Lett. {\bf
  78},  2275  (1997).

\bibitem{VaidmanL94}
L. Vaidman, Phys. Rev. A {\bf 49},  1473  (1994).

\bibitem{BraunsteinSL98}
S.~L. Braunstein and H.~J. Kimble, Phys. Rev. Lett. {\bf 80},  869  (1998).

\bibitem{FurusawaA98}
A. Furusawa, J.~L. Sorensen, S.~L. Braunstein, C.~A. Fuchs, H.~J. Kimble, and
  E.~S. Polzik, Science {\bf 282},  706  (1998).

\bibitem{LloydS98}
S. Lloyd and J.-J.~E. Slotine, Phys. Rev. Lett. {\bf 80},  4088  (1998).

\bibitem{BraunsteinSL98a}
S.~L. Braunstein, Phys. Rev. Lett. {\bf 80},  4084  (1998).

\bibitem{LloydS99}
S. Lloyd and S.~L. Braunstein, Phys. Rev. Lett. {\bf 82},  1784  (1999).

\bibitem{LeuchsG99}
G. Leuchs, T.~C. Ralph, C. Silberhorn, and N. Korolkova, J. Mod. Opt. {\bf 46},
   1927  (1999).

\bibitem{JoobeurA96}
A. Joobeur, B.~E.~A. Saleh, T.~S. Larchuk, and M.~C. Teich, Phys. Rev. A {\bf
  53},  4360  (1996).

\bibitem{KellerTE97}
T.~E. Keller and M.~H. Rubin, Phys. Rev. A {\bf 56},  1534  (1997).

\bibitem{HasegawaA89}
A. Hasegawa, {\em Optical Solitons in Fibers} (Springer-Verlag, Berlin, 1989).

\bibitem{ReynaudF92}
F. Reynaud and A. Barthelemy, NATO ASI Series E {\bf 214},  319  (1992).

\bibitem{ZakharovVE72}
V.~E. Zakharov and A.~B. Shabat, Soviet Physics - JETP {\bf 34},  62  (1972).

\bibitem{SatsumaJ74}
J. Satsuma and N. Yajima, Prog. Theor. Phys. Suppl. {\bf 55},  284  (1974).

\bibitem{MollenauerLF80}
L.~F. Mollenauer, R.~H. Stolen, and J.~P. Gordon, Phys. Rev. Lett. {\bf 45},
  1095  (1980).

\bibitem{BarthelemyA85}
A. Barthelemy, S. Maneuf, and C. Froehly, Opt. Commun. {\bf 55},  201  (1985).

\bibitem{ManeufS88}
S. Maneuf and F. Reynaud, Opt. Commun. {\bf 66},  325  (1988).

\bibitem{LaiY89}
Y. Lai and H.~A. Haus, Phys. Rev. A {\bf 40},  844  (1989).

\bibitem{MecozziA98}
A. Mecozzi and P. Kumar, Quantum Semiclass. Opt. {\bf 10},  L21  (1998).

\bibitem{SchmidtE99}
E. Schmidt, L. Kn{\"o}ll, and D.-G. Welsch, Phys. Rev. A {\bf 59},  2442
  (1999).

\bibitem{SpaelterS98}
S. Sp{\"a}lter, N. Korolkova, F. K{\"o}nig, A. Sizmann, and G. Leuchs, Phys.
  Rev. Lett. {\bf 81},  786  (1998).

\bibitem{MecozziA97}
A. Mecozzi and P. Kumar, Opt. Commun. {\bf 22},  1232  (1997).

\bibitem{LevandovskyD99}
D. Levandovsky, M. Vasilyev, and P. Kumar, Opt. Lett. {\bf 24},  43  (1999).

\bibitem{SchmidtE2000}
E. Schmidt, L. Kn{\"o}ll, and D.-G. Welsch, Quantum noise of damped
  $N$-solitons, in press: Opt. Commun., 2000.

\bibitem{FribergSR96}
S.~R. Friberg, S. Machida, M.~J. Werner, A. Levanon, and T. Mukai, Phys. Rev.
  Lett. {\bf 77},  3775  (1996).

\bibitem{SpaelterS97}
S. Sp{\"a}lter, M. Burk, U. Str{\"o}ssner, M. B{\"o}hm, A. Sizmann, and G.
  Leuchs, Europhys. Lett. {\bf 38},  335  (1997).

\end{thebibliography}
%
%

\newpage
\multicolsend

\vspace*{0.5cm}

\newpage

\vspace*{0.5cm}



\end{document}